\documentclass[12pt]{iopart}
\usepackage[dvips]{graphicx}
\usepackage{amssymb,amsopn, verbatim,latexsym}
\usepackage[bookmarks,dvips,colorlinks, pdfauthor={Martin Horvat}, 
            pdftitle={Dynamical approach to chains of scatterers}]{hyperref}
 
\begin{document}

\frenchspacing

\def\picbox#1#2{\fbox{\vbox to#2{\hbox to#1{}}}}
\def\bra#1{\langle#1|}
\def\ket#1{|#1\rangle}
\def\braket#1#2{\langle#1|#2\rangle}
\def\ave#1{\left\langle #1 \right\rangle}
\def\parc#1#2{\frac{\partial #1}{\partial #2}}
\def\pa{\partial} 
\def\scalar#1#2{\langle#1|#2\rangle}
\def\eps{\varepsilon}
\def\mean#1{\overline{#1}}

\def\bC{{\mathbb{C}}}
\def\bZ{{\mathbb{Z}}}
\def\bR{{\mathbb{R}}}
\def\bN{{\mathbb{N}}}

\def\dd{{\rm d}}
\def\id{{\rm 1}}
\def\ii{{\rm i}}
\def\rL{{\rm L}}
\def\rR{{\rm R}}

\def\asin{\operatorname{arcsin}}
\def\acos{\operatorname{arccos}}
\def\asinh{\operatorname{arcsinh}}
\def\acosh{\operatorname{arccosh}}
\def\atan{\operatorname{arctan}}
\def\atanh{\operatorname{arctanh}}
\def\tr{\operatorname{tr}}
\def\vol{\operatorname{vol}}
\def\rot{\operatorname{rot}}
\def\grad{\operatorname{grad}}
\def\diag{\operatorname{diag}}
\def\card{\operatorname{card}}
\def\const{\operatorname{const}}

\def\beq{\begin{equation}}
\def\eeq{\end{equation}}
\def\beqa{\begin{eqnarray}}
\def\eeqa{\end{eqnarray}}
\def\beqaa{\begin{eqnarray*}}
\def\eeqaa{\end{eqnarray*}}

\def\mymat#1#2#3#4{%
{\left[\begin{array}{cc} #1 & #2 \cr #3 & #4\end{array}\right]}%
}
\def\myvec#1#2{%
{\left[\begin{array}{c} #1 \cr #2 \end{array}\right]}%
}
\def\rH{{\rm H}}
\def\rNo{N_{\rm o}}
\def\rNc{N_{\rm c}}
\def\roo{{\rm oo}}

\title{Dynamical approach to chains of scatterers}

\author{Martin Horvat and Toma\v z Prosen}

\address{Physics Department, Faculty of Mathematics and Physics, 
University of Ljubljana, Slovenia}

\eads{\mailto{martin.horvat@fmf.uni-lj.si}, 
      \mailto{tomaz.prosen@fmf.uni-lj.si}}  
\begin{abstract}
Linear chains of quantum scatterers are studied in the process of lengthening, which is treated and analysed as a discrete dynamical system defined over the manifold of scattering matrices. Elementary properties of such dynamics relate the transport through the chain to the spectral properties of individual scatterers.  For a single-scattering channel case some new light is shed on known transport properties of disordered and noisy chains, whereas translationally invariant case can be studied analytically in terms of a simple deterministic dynamical map. The many-channel case was studied numerically by examining the statistical properties of scatterers that correspond to a certain type of transport of the chain i.e. ballistic or (partially) localised.
\end{abstract}

\submitto{\JPA}


\section{Introduction}

Linear chains of scatterers represent a useful model of macroscopic structures like multi-layered structures, real-life wires, nano-tubes etc. This is the reason that the scattering chains have attracted a lot of scientific attention from the early 1980s to the present time. Past studies mainly focused on the chain of randomly chosen scatterers and its average properties. For good reviews on this topic see \cite{erdos:adp:82, been:rev_mod:97, nakamura:ch_sol:97,kramer:rep_prog:93}, where the main approach of theoretical analysis is the transfer matrix formalism \cite{newton:book:02}. Some other interesting articles discussing linear chain of scatterers with some disorder using the same approach are \cite{cahay:phys_rev:88, abrahams:jpc:80, andereck:jpc:80, kirkman:jpc:84, langley:jsv:96}. There were also attempts to understand the chains using purely scattering matrix approach, although they have only partially used its advantages. For original references along these lines see Refs. \cite{anderson:phys_rev:80, anderson:phys_rev:82,ko:phys_rev:88}.
A considerable breakthrough in understanding of localisation in disordered wires was made using the Dorokhov-Mello-Pereyra-Kumar (DMPK) scaling equation \cite{dorokhov:jetpl:82, mello:ann_phys:88}, which gives the scaling of the distribution of transmission-like-quantities of individual modes with the chain length. 
More recently there was some renewed interest in scattering formalism, for example in the stability analysis of scattering matrix merging procedures \cite{mayer:pre:99}, and  the random walk in the scattering chains \cite{cwilich:nano:02}.\par
\begin{figure}[!htb]
  \centering
  \includegraphics[width=10cm]{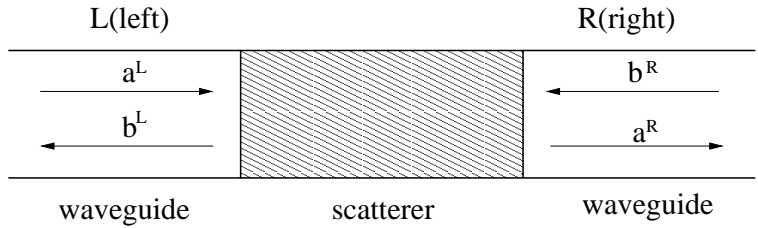}
\caption{Schematic picture of a single scatterer representing a basic
cell of the chain in discussion.}
\label{pic:cell}
\end{figure}

In this paper we examine linear chains of abstract quantum scatterers. 
The scatterers on the chain can be any conservative open quantum systems generating wave-dynamics, and having two identical waveguides attached on the left/right side which connect to the previous/next scatterer (see figure \ref{pic:cell}). We shall leave out from discussion all geometric parameters of the systems and the wave-guides and work purely algebraically to make results as general as possible.
We are interested in general {\it on-shell transport properties} of these chains in the process of lengthening, namely in what we shall refer to as {\it the dynamical approach to scattering}. We work within the scattering matrix formalism in order to avoid divergences present in e.g. transfer matrix formalism, and to take advantage of the compactness of the phase space manifold of scattering matrices. The presented dynamical approach enables intuitively clear insight into the finite and infinite chains. In the following sections, besides discussing dynamical properties of lengthening, we also derive some interesting transport properties of the chain in three different cases: translationally invariant chain, chain with weak disorder and chain with strong disorder. Some of our results can be understood as re-derivation of known transport properties of one-dimensional lattices from a simple new dynamical perspective, in particular for the case of single-channel scatterers. However, in case of multi-channel scatterers we report some new - numerical - results on the scaling of Haar measures of scattering matrices corresponding to ballistic and localised dynamics.
 
\section{Scattering matrix formalism}

In this section we give a short sketch of scattering formalism in quantum chains,
but the reader can see \cite{londergan:book:99} for details. The stationary Schr\" odinger equation for the open quantum problem composed of the scatterer and two infinite wave-guides is written symbolically as
\beq
  \hat H  \ket{\psi} = E \ket{\psi}\>,
\eeq
where $\hat H$ represents the Hamiltonian of the system. The wave functions in the left $\ket{\psi_\rL}$ and in the right waveguide $\ket{\psi_\rR}$ are expanded in the channel basis $\{\ket{e^{\pm}_n}\}_{n=1}^d$ 
\beq
  \ket{\psi_{\rL,\rR}}
  =
  \sum_{n=1}^d a_n^{\rL,\rR} \ket{e^+_n} + b_n^{\rL,\rR} \ket{e^-_n}\>,
  \label{eq:wave}
\eeq
where superscript $+$ and $-$ corresponds to the phase propagation from left-to-right and right-to-left, respectively. The number of basis mode functions involved in the expansion, denoted by $d$, is also called the {\it number of (scattering) channels}. The smoothness of the wave function on the boundary between the wave-guides and the scatterer gives the condition which connects the wave functions on both sides of the scatterer. The connection between the two sides expressed in vectors of expansion coefficients $a^{\rL, \rR}=\{a_n^{\rL, \rR}\}_{n=1}^d$ and $b^{\rL, \rR} = \{ b_n^{\rL,\rR} \}_{n=1}^d$ can be given in the form of a {\it scattering matrix} $S$ or a {\it transfer matrix} $T$:
\beqa
  S \psi_{\rm in} = \psi_{\rm out}\>,
  \qquad
  \psi_{\rm in}  = \myvec{a^\rL}{b^\rR},\quad
  \psi_{\rm out} = \myvec{b^\rL}{a^\rR}\>,
  \label{eq:S_def} \\
  T\psi_\rL = \psi_\rR\>,\qquad
  \psi_{\rm L}  = \myvec{a^\rL}{b^\rL}\>,\quad
  \psi_{\rm R} = \myvec{a^\rR}{b^\rR}\>,
 \label{eq:T_def}
\eeqa
with superscripts L and R corresponding to the left and the right side of the scatterer. Following the definitions (\ref{eq:S_def}) and (\ref{eq:T_def}) it is convenient to express the scattering matrix $S$ and the transfer matrix $T$ as block matrices
\beq
  S = \mymat{r^\rL}{t^\rR}{t^\rL}{r^\rR}\>,\quad
  T = \mymat{x_1}{x_2}{x_3}{x_4}\>,
  \label{eq:ST_form}
\eeq
where $r^{\rL,\rR}$ represent the reflection matrices and $t^{\rL,\rR}$ the transmission matrices for incidence of the left and right side of the scatterer. The matrices of this form are sometimes called {\it two-way (port) scattering and transfer matrices}, because they connect two distant openings of the scatterer. Purely from definitions (\ref{eq:S_def}) and (\ref{eq:T_def}) we obtain explicit relations between the scattering matrix and the transfer matrix
\beq
  \fl\hspace{1cm}
  T[S] = \mymat{t^\rL - r^\rR {t^\rR}^{-1} r^\rL}{r^\rR {t^\rR}^{-1}} 
               {-{t^\rR}^{-1} r^\rL}{{t^\rR}^{-1}}\>,\quad
  S[T] = \mymat{-x_4^{-1} x_3}{x_4^{-1}}
               {x_1 - x_2 x_4^{-1} x_3}{x_2 x_4^{-1}}\>.
\label{eq:ST_relation}
\eeq
From conservation of probability currents it follows (see e.g. \cite{newton:book:02})
that the matrices $S$ and $T$ fulfill the following relations
\beq
  S^\dag S = S S^\dag = \id\>,\quad  T^\dag K T = T K T^\dag = K\>,\quad 
  K = \mymat{\id}{0}{0}{-\id}\>,
  \label{eq:symm}
\eeq
implying that the scattering matrix $S$ is unitary, $S\in U(2d)$, and the transfer matrix $T$ is hyperbolic, $T\in U(d,d)$. Then by introducing {\it transport probability matrices} corresponding to the reflection $\Pi_x$ and transmission $\Sigma_x$
\beq
  \Pi_x = {r^x}^\dag r^x\>,\quad
  \Sigma_x = {t^x}^\dag t^x\>,\quad \Pi_x + \Sigma_x = \id\>, \qquad 
  x=\rL,\rR\>. 
\eeq
we define the two basic measures of transport of the scatterer : (i) the {\it average transmission (reflection) probability} $\cal T$ ($\cal R$) as
\beq
  {\cal R} = \ave{\Pi_x}\>,\quad
  {\cal T} = \ave{\Sigma_x} =  1 - {\cal R}\>,\qquad 
  x = \rL,\rR\>. 
\label{eq::qch:ave_transp}
\eeq
and (ii) the {\it standard deviation of the transmission (reflection) probability} $\sigma^2_T$ ($\sigma^2_R$) as 
\beq
  \sigma^2_{\rm R} = \frac{1}{d+1}\left[\ave{\Pi_x^2} - {\cal R}^2\right] = 
  \sigma^2_{\rm T} = \frac{1}{d+1}\left[\ave{\Sigma_x^2} - {\cal T}^2\right]\>,
  \qquad x = \rL,\rR\>, 
\label{eq::qch:dev_transp}
\eeq
where we have used the symbol $\ave{\bullet}=\frac{1}{d} \tr\{\bullet\}$. We see that only two quantities can be independent and so we use either $({\cal R}, \sigma_{\rm R})$ or $({\cal T}, \sigma_{\rm T})$ as a measure of transport.\par
\begin{figure}[!htb]
  \centering
  \includegraphics[width=12cm]{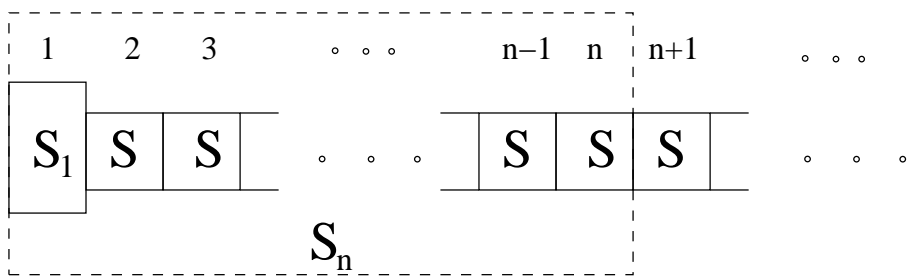}
\caption{Schematic picture of a linear chain of scatterers. It is build by recursive lengthening of the initial chain $S_1$ with scatterers $S$. Scatterer $S$ can be taken as fixed or $n$-dependent.}
\label{pic:chain}
\end{figure}

We build the chain of scatterers recursively by connecting additional scatterers to one end of the chain. This procedure is illustrated in figure \ref{pic:chain}. The chain's length is measured in units of the number of scatterers composing the chain. An existing chain of length $n$ with the scattering matrix $S_n$ is extended with an additional elementary scatterer described by a {\it generating scattering matrix} $S$, forming a chain of length $n+1$, and with the scattering matrix $S_{n+1}$:
\beq
  S = \mymat{r^\rL}{t^\rR}{t^\rL}{r^\rR}\>,\quad
  S_n = \mymat{r_n^\rL}{t_n^\rR}{t_n^\rL}{r_n^\rR}\>.
\eeq
The recurrence relation for the scattering matrices of the chain then reads
\beq
  S_{n+1} = S_n \odot S\>,
  \label{eq:S_map}
\eeq
where we introduced a binary operation $\odot$ representing concatenation of the scattering matrices. It is explicitly written as
\beqa
  r^\rL_{n+1} &=& r_n^\rL + 
           t_n^\rR r^\rL L^{-1} t_n^\rL\>,\quad 
  t^\rL_{n+1} = t^\rL L^{-1}t_n^\rL\>,\\
  r^\rR_{n+1} &=& r^\rR + 
           t^\rL r_n^\rR {L'}^{-1} t^\rR\>,\quad 
  t^\rR_{n+1} = t_n^\rR {L'}^{-1}t^\rR\>,
  \label{eq:S_map_explicit}
\eeqa
with matrix expressions $L = \id - r_n^\rR r^\rL$ and $L' = \id - r^\rL r_n^\rR$. Note that unitary scattering matrices $U(2d)$ form a group \cite{scott:book:87} with the operation $\odot$. We think of the recurrence (\ref{eq:S_map}) as a discrete dynamical system defined over the space of scattering matrices. We study two types of chain-generation: either we take a fixed (static) generating scattering matrix $S$, or $S$ is taken randomly during the growth of the chain. The iteration (\ref{eq:S_map}) can be written in the transfer matrix formalism in terms of matrix products as
\beq
  T_{n+1} = T T_n\>,
  \label{eq:T_map}
\eeq
where $T = T[S]$ is the {\it generating transfer matrix} and $T_n = T[S_n]$ is the transfer matrix of the chain of length $n$. But as can be read from the relations (\ref{eq:symm}), matrix elements of $T_n$ are not bounded in size. This implies that the map (\ref{eq:T_map}) is generally unstable and so numerically of limited use. 
\section{Chain of single-channel scatterers}
Here we discuss chains of linearly connected single-channel ($d=1$) scatterers. We describe the chain and the individual scatterers in terms of $2\times 2$ unitary scattering matrices $S$ parametrised as
\beq
  S(A,\alpha^\rL,\beta^\rL,\beta^\rR) = 
  \left[
    \begin{array}{cc} 
      A e^{\ii \alpha^\rL} & B e^{\ii \beta^\rR} \cr
      B e^{\ii \beta^\rL} & -A e^{\ii (\beta^\rL + \beta^\rR - \alpha^\rL)}
    \end{array}
  \right]\>,\quad A = \sqrt{1-B^2}\>,
  \label{eq:S_1dparam}
\eeq
where $A$ and $B$ are square roots of reflection and transmission intensity, respectively, $\alpha^\rL$ is the reflection phase, and $\beta^{\rR,\rL}$ are the transmission phases. Therefore any unitary $2\times 2$ matrix represents a physically legitimate scattering matrix. Note that ${\cal T}[S] = B^2$, ${\cal R}[S] = A^2$ and $\sigma_R=\sigma_T=0$. In introduced parametrisation (\ref{eq:S_1dparam}) the scattering matrix $S_n$ of the chain of length $n$ and the generating scattering matrix $S$ read 
\beq
  S_n = S(A_n, \alpha_n^\rL, \beta_n^\rL, \beta_n^\rR)\>,\qquad 
  S = S(A, \alpha^\rL,  \beta^\rL, \beta^\rR)\>.
  \label{eq:S_1dparam_mat}
\eeq
The recurrence relation/map for the chain generation (\ref{eq:S_map}) can now be written out explicitly.  We discuss two cases of chain generation. In the first case the chain is translationally invariant with fixed $S$ and in the second case $S$ is chosen randomly at each iteration step creating a random (disordered) chain. The appropriate explicit form of the map differs a bit for the two cases.
\subsection{Static generating scattering matrix}
\begin{figure}[!htb]
  \centering
  \includegraphics[width=7cm]{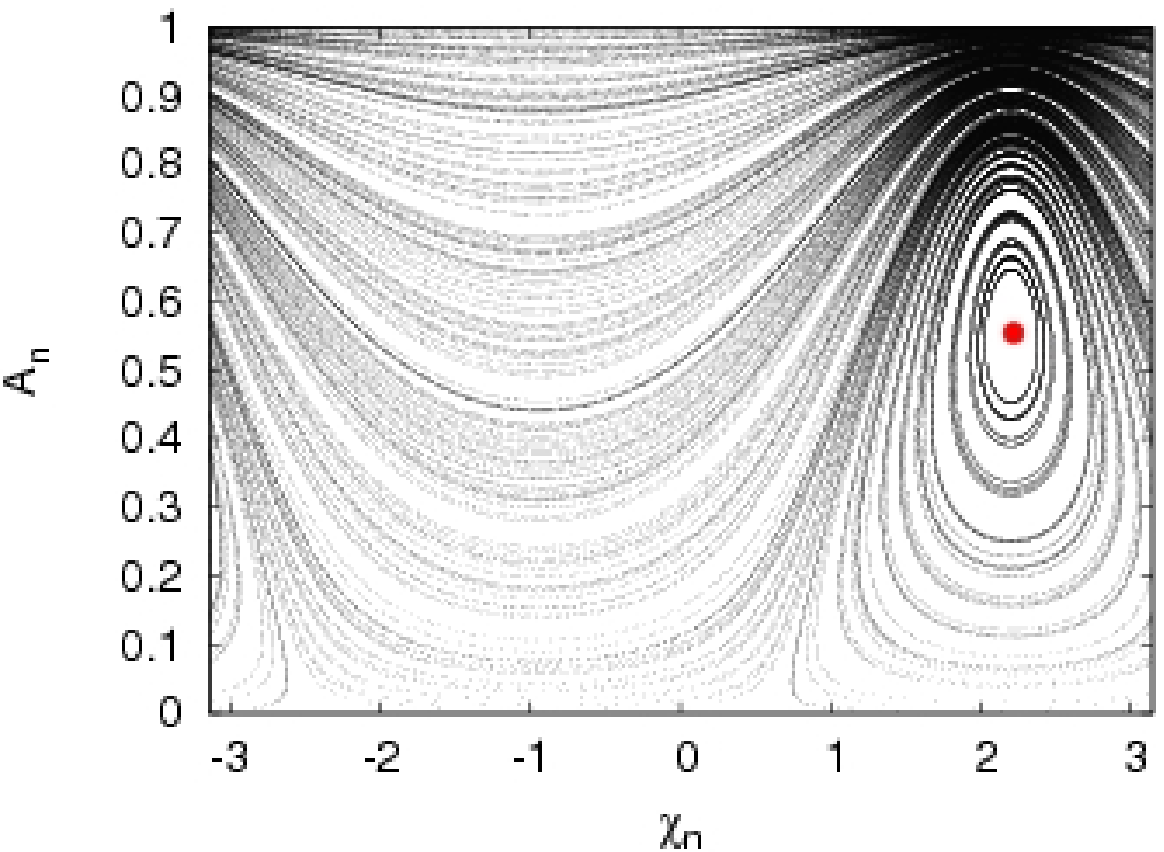}\hskip2pt%
  \includegraphics[width=7cm]{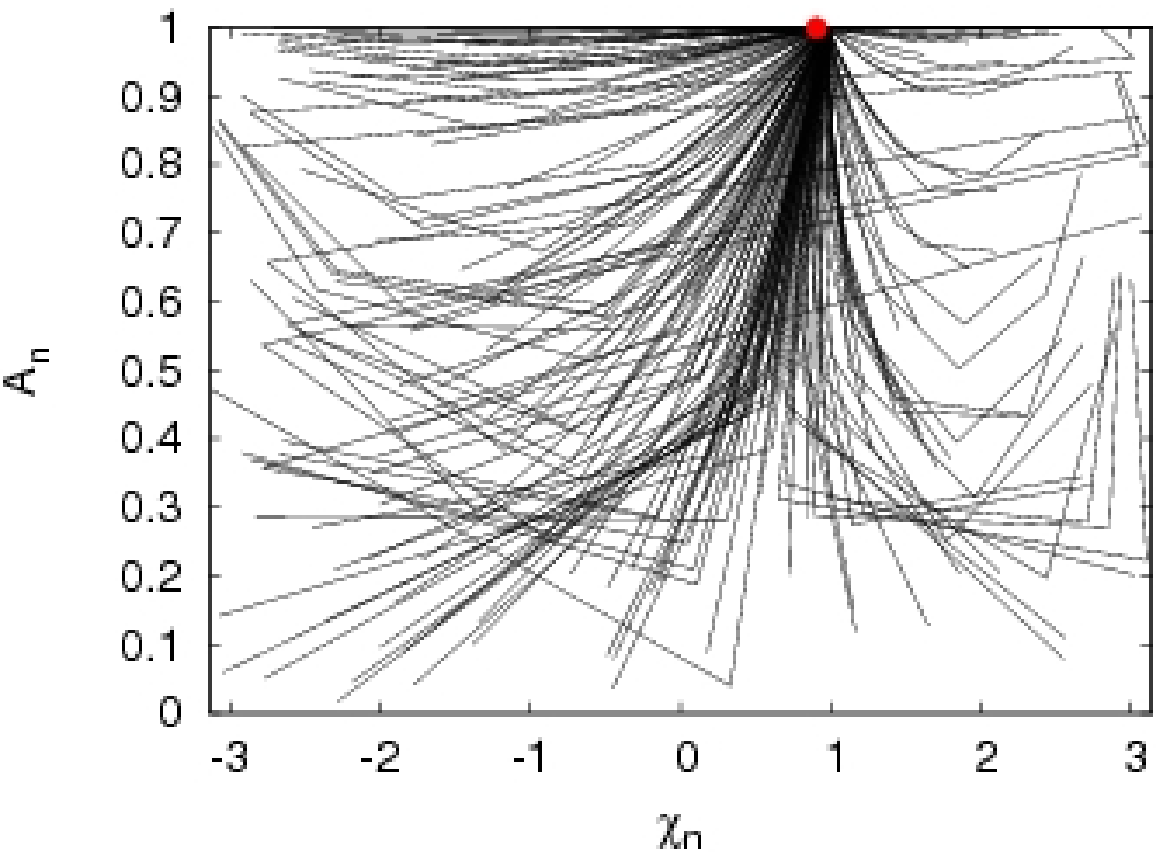}
  \hbox to16cm{\hfil (a) \hfil\hfil (b) \hfil}
\caption{The phase space portraits of the dynamical system (\ref{eq:static_A},\ref{eq:static_chi}) generated at parameters $A = 0.5$, $\lambda = 0.628319$ (a) and $A = 0.5$, $\lambda = 0.314159$ (b). In (b) we connect
consequent phase space points in order to give some rough impression on the direction of dynamics.
Positions of fixed points are indicated by red circular dots.
}
\label{pic:portrait1}
\end{figure}

Here we consider an initial scatterer with scattering matrix $S_1$ and the generating matrix $S$ which is constant along the chain. The infinite chain represents a quantum particle in a one-dimensional periodic structure on a half-line with a given initial condition. For a scalar periodic potential the problem is discussed in a standard literature on quantum mechanics using the transfer matrix approach, see e.g.\cite{cohen-tannoudji:book:06} p.367. In a related problem on a doubly-infinite line, the spectrum has a form of energy bands and the standard Bloch theorem applies. In such a case, the explicit form of the recurrence relation for the chain generation (\ref{eq:S_map}), by using a new variable $\chi_n =\beta_n^\rL + \beta_n^\rR - \alpha_n^\rL + \alpha $, reads
\beqa
  A_{n+1} &=& \sqrt{\frac{A_n^2 + A^2 + 2A_nA\cos\chi_n}
                     {1 + 2A_nA\cos\chi_n + (A_n A)^2}}\>,
 \label{eq:static_A}\\
  \chi_{n+1} &=& \chi_n + 
  2\lambda + \arg \{(1 + A_n A e^{-\ii \chi_n})(A_n + A e^{-\ii \chi_n})\}\>,
  \label{eq:static_chi}\\
  \beta_{n+1}^\rL	&=& \beta_n^\rL  + 
  \beta^\rL + \arg\{(1 + A_n A e^{-\ii \chi_n})\}\>,\\
  \beta_{n+1}^\rR	&=& \beta_n^\rR  + 
  \beta^\rR + \arg\{(1 + A_n A e^{-\ii \chi_n})\}\>.
\eeqa
where we have introduced a parameter $\lambda = (\beta^\rL + \beta^\rR)/2$ of the generating scattering matrix. The relevant parts of this system of discrete equations are expressions (\ref{eq:static_A}) and (\ref{eq:static_chi}) representing an autonomous two-dimensional dynamical system. Note that this dynamical system would not change if we restrict ourselves to the case of time-reversal invariant (symmetric) S matrices for which $\beta^{\rm L}_n \equiv \beta^{\rm R}_n$. Depending on parameters of the generating matrix $S$, namely $A$ and $\lambda$, we distinguish two topologically different types of dynamics: (i) quasi-periodic or {\it ballistic} with an elliptic fixed-point and (ii) convergence to a single attractive fixed-point, called {\it localisation}. The phase-space portraits for both types of dynamics are plotted in figure \ref{pic:portrait1}.
The study of our two-dimensional dynamical system shows that the type of dynamics depends on the sign of the following discriminant
\beq
  D = A^2 - \sin^2\lambda \>,
\eeq
yielding for $D < 0$ the ballistic and for $D > 0$ the localised dynamics. The expression $D$ is a discriminant in the eigenvalue problem of the transfer matrix $T[S]$ $(\det (T[S] - \kappa\, \id) = 0)$:
\beq
\kappa_{1,2} = \frac{e^{\pm \ii \frac{\beta^\rL - \beta^\rR}{2}}}{\sqrt{1-A^2}} \left( \cos\lambda \pm \sqrt{D}\right)
\eeq
and it is easy to see that, suitably ordering $\kappa_{1,2}$,
\beq
  D<0 \Rightarrow \kappa_1=\kappa_2^*{\rm ~~and~~} |\kappa_{1,2}|=1\>,\qquad 
  D>0 \Rightarrow |\kappa_1|< 1 < |\kappa_2|\>.
\eeq
We can conclude that a growing chain in the transfer matrix formalism (\ref{eq:T_map}) for $D < 0$ will be numerically stable and for $D > 0$ some matrix elements of the transfer matrix $T_n[S_n]$ will diverge.\par
We note that for the corresponding problem on an  infinite line the ballistic case $D < 0$, and the localised case $D > 0$, correspond to physical energy being positioned inside the {\em (Bloch) energy band}, or in the {\em gap}, 
respectively.\par
In case of ballistic dynamics, $D < 0$, we find a unique elliptic fixed point $(A_{\rm e},\chi_{\rm e})$:
\beqa
  A_{\rm e} &=& u - \sqrt{u^2 -1}\>,\quad u =\frac{\sin\lambda}{A}\>, \\
  \chi_{\rm e} &=& \lambda + {\pi\over 2} + m\pi \>,
\eeqa
where $m \in \bZ$ is such that $\chi_{\rm e} \notin (-\pi/2, \pi/2)$. Further, phase-space portrait of ballistic dynamics indicates an existence of an additional integral of motion $F$ of dynamics (\ref{eq:static_A}, \ref{eq:static_chi}), which can be derived using  the correspondence between the scattering and the transfer matrices given by (\ref{eq:ST_relation}):
\beq
  F(A_n,\chi_n) = \frac{\sin\lambda + A_n A \sin(\lambda -\chi_n)}{1-A_n^2}\>.
  \label{eq:integral}
\eeq
By considering the last expression we conclude that the scattering matrix of the chain evolves on a 
one-dimensional manifold. 

In the case of localised dynamics, $D > 0$, the whole phase-space converges to a single fixed point attractor $(A_{\rm a}, \chi_{\rm a})$:
\beqa
  A_{\rm a} &=&1\>,\\
  \chi_{\rm a}  &=& \lambda +\arg\{\ii u + \sqrt{1-u^2}\}\>,\quad
  u = \frac{\sin \lambda}{A}\>.
\eeqa
Hence, the chain converges with increasing length to a state of zero transmission (perfect reflection). The convergence is exponential. This can be best seen by locally expanding the dynamics around the fixed-point in variables $B = \sqrt{1-A^2}$ and $\chi_n = \chi_a + \delta \chi_n$ and so we obtain
\beq
  B_n,\; \delta \chi_n \sim |\kappa_1|^{n}\>.
\eeq
Note that the norms of all matrix elements converge to limiting values, but the phases do not, with the exception of $r_\rL$. Similar behaviour was encountered for multi-channel scattering chains.
\subsection{Noisy generating matrix}
Here we discuss disordered chains that are composed of randomly chosen scatterers. Such chains of scatterers are known as {\it random chains} or {\it random wires}. The scattering matrix $S_n$ of a chain of length $n$ is constructed from its initial state $S_1$ by merging with generating matrices $S$ that are randomly chosen on each step of lengthening. To simplify the discussion we define three disjoint sets of generating scattering matrices:
\beqa
  {\cal M}_{\rm b} &=& \{S \in U(2) : D < 0\}\>,\\
  {\cal M}_{\rm l} &=& \{S \in U(2) : D > 0\}\>,\\
  {\cal M}_{\rm m} &=& \{S \in U(2) : D = 0\}\>.
\eeqa
We call ${\cal M}_{\rm b}$ the set of {\it ballistic matrices}, ${\cal M}_{\rm l}$, the set of {\it localised matrices} and ${\cal M}_{\rm m}$ the set of {\it marginal matrices}. In order to measure the volume of introduced sets we use a uniquely defined invariant measure over unitary matrices $\mu_\rH$, called {\it the Haar measure} \cite{reichl:book:04}, and normalised so that $\mu_\rH (U(2d)) = 1$. 
The Haar measure of the marginal matrices $\mu_H({\cal M}_{\rm m})$ is obviously zero thereby making this set uninteresting for our general discussion. The measures of the other two sets are the following:
\beqa
%
  \mu_\rH({\cal M}_{\rm b}) &=& 1 -\mu_\rH({\cal M}_{\rm l}) = \frac{1}{2}\>,
\eeqa
%
%
It is important to think about the role of these sets in the construction of the chains. Let us construct a chain of length $n$ in which we use $m(n)$ localised generating matrices from ${\cal M}_{\rm l}$. It is evident that in case the ratio $m(n)/n$ in the limit $n\to\infty$ is finite then the transmission of the chain converges exponentially towards zero. In the opposite case, when the chain is constructed mostly of ballistic generating matrices from ${\cal M}_{\rm b}$, the transmission can in the worst case decrease linearly with the length of the chain.\par
\begin{figure}[!htb]
  \centering
  \includegraphics[width=7cm]{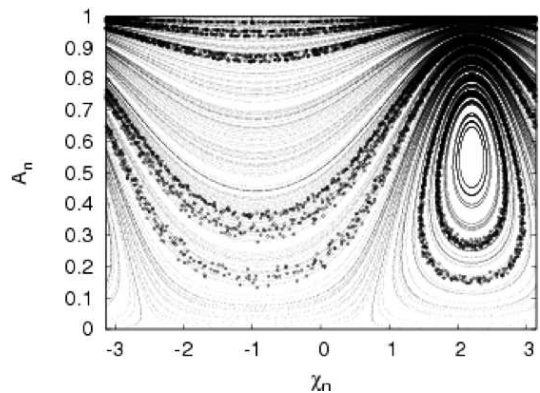}
\caption{The phase space portrait in the ballistic regime for parameters $A_0 = 0.5$ and $\lambda_0 = 0.628319$ without the noise (small points) and with few sample trajectories obtained in the presence of noise in the generating matrix $S$ (large points). The latter is defined by $A = A_0  + \epsilon \zeta$, $\lambda = \lambda_0 + \epsilon\zeta$, where $\epsilon = 0.001$ and $\zeta \in [-1, 1]$ is uniformly distributed stochastic variable.}
\label{pic:portrait2}
\end{figure}

Now we consider chains of single-channel scatterers in which the generating scattering matrix explicitly depends on the position in the chain. In the parametrisation (\ref{eq:S_1dparam_mat}), the dynamics of the chain's scattering matrix in the process of lengthening is determined by the system of discrete equations
\beqa
  B_{n+1} 
  &=& \frac{B_n B}{\sqrt{1 + 2A_nA\cos(\phi_n + \alpha^\rL) + (A_n A)^2}}\>,
  \label{eq:dyn_B}\\
  \phi_{n+1} &=& \phi_n + 2\lambda + 
  \arg \{(1 + A_n A e^{-\ii (\phi_n + \alpha^\rL)})
         (A_n + A e^{-\ii (\phi_n + \alpha^\rL)})\}\>,
  \label{eq:dyn_phase}\\
  \beta_{n+1}^\rL	&=&  \beta_n^\rL  + 
  \beta^\rL + \arg\{(1 + A_n A e^{-\ii (\phi_n + \alpha^\rL)})\}\>,\\
  \beta_{n+1}^\rR	&=& \beta_n^\rR  + 
  \beta^\rR + \arg\{(1 + A_n A e^{-\ii (\phi_n + \alpha^\rL)})\}\>,
\eeqa
where we introduce additional variable $\phi_n = \beta_n^\rL + \beta_n^\rR - \alpha_n^\rL = \chi_n - \alpha^\rL $. This four dimensional dynamical system can be reduced to a two dimensional one. It is described by the transmission intensity $B_n$, phase $\phi_n$ and evolution equations (\ref{eq:dyn_B}) and (\ref{eq:dyn_phase}). Each iteration step is controlled by three parameters $A=A(n)$, $\lambda=\lambda(n)$ and $\alpha^\rL=\alpha^\rL(n)$. This is one parameter more in comparison to the static case of constant generating matrix.
Just for illustration, we show in figure \ref{pic:portrait2} an example of the chain generated by a noisy scattering matrix $S$ chosen deep enough in the ballistic set ${\cal M}_{\rm b}$. We see that the presence of a small noise does not strongly deform the trajectory, namely, {\em it does not produce localization}. It just induces a small variation around unperturbed trajectory given by the integral of motion (\ref{eq:integral}) for the average generating scattering matrix $\ave{S}$. Note a similarity with the Kolmogorov-Arnold-Moser type stability in Hamiltonian dynamics.\par
\begin{figure}[!htb]
  \centering
  \includegraphics[width=7cm]{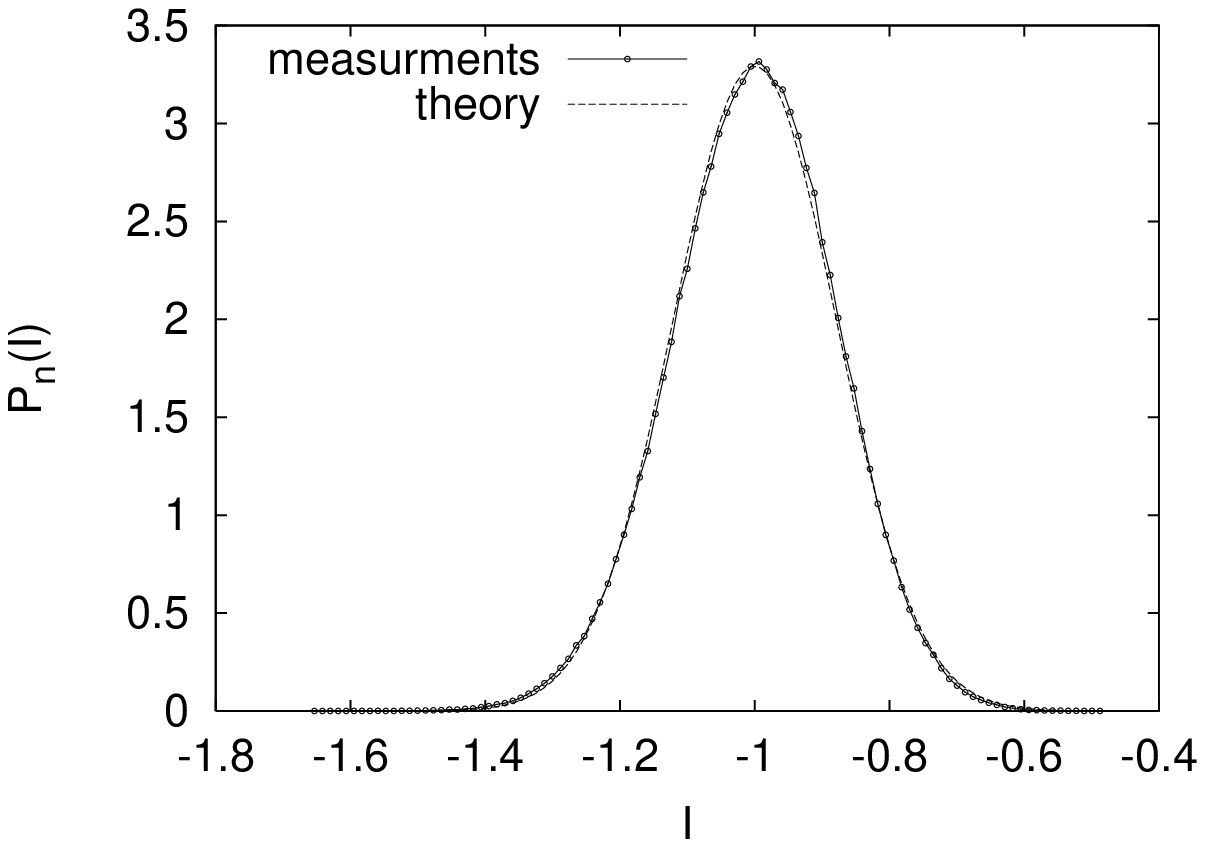}%
  \includegraphics[width=7cm]{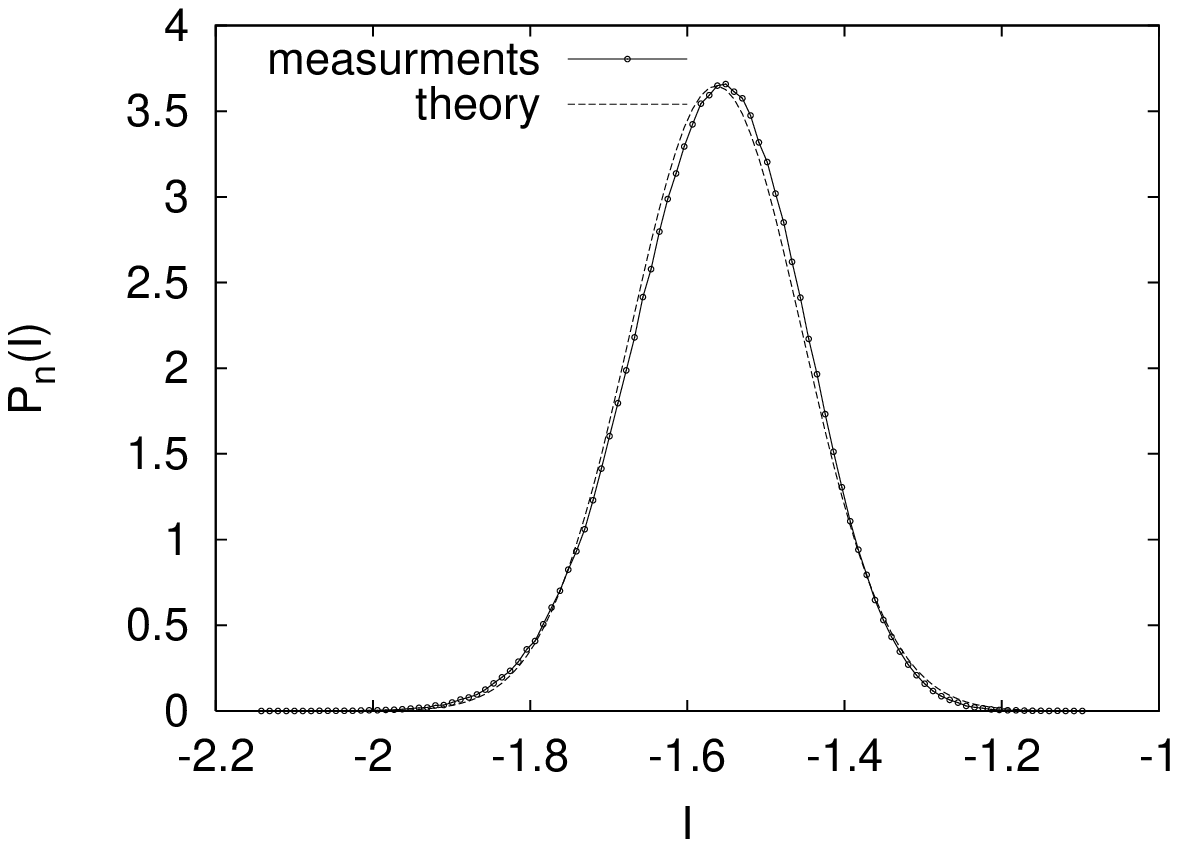}
  \hbox to16cm{\hfil (a) \hfil\hfil (b) \hfil}
\caption{Theoretical and measured distribution of the average decay
rate $I_n$ at $n = 100$ for two examples of random chains, where $\lambda = \pi/10$, $B \in [0, 1]$ uniformly distributed. Variable $\alpha^{\rL}$ was uniform over $[0, 2\pi]$ in case (a) and over $[0.5,0.7]$ in case (b). The theoretical curves are given with $\mean{I}=-1$, $\sigma_I^2 = 1.4674$ in the case (a) and with $\mean{I}=-1.56325$, $\sigma_I^2=1.19556$ in the case (b).}
\label{pic:distr_I}
\end{figure}

Let us now discuss a general case of random chains with {\em strong disorder}. These chains are built using generating matrices from some set $\cal A$ intersecting the set of localised matrices ${\cal M}_{\rm l}$ so that $\mu_\rH ( {\cal A} \cap {\cal M}_{\rm l})> 0$. We are mainly interested in the transmission properties of asymptotically long chains. Intuitively we expect that the transmission will converge exponentially towards zero exhibiting {\it  exponential localisation}. In the limit of long chains the transmission is very small and we can replace the exact evolution with the following approximation:
\beqa
  B_{n+1}    &=& B_n f(B(n),\phi_n + \alpha(n))\>,
  \label{eq:approx_B}\\
  \phi_{n+1} &=& \phi_n + 2\lambda(n) + 
               2 \arg \{ 1 + \sqrt{1-B(n)^2} e^{-\ii (\phi_n + \alpha(n))}\}\>,
  \label{eq:approx_phi}\\
  f(x,y)    &=& \frac{x}{\sqrt{1 + 2\sqrt{1-x^2}\cos(y) + (1-x^2)}}\>,
\eeqa
where $B(n)$, $\alpha(n)$ and $\lambda(n)$ are parameters of the generation scattering matrix $S\in {\cal A}$ in $n$-th iteration step. The matrix $S$ is picked randomly at each iteration step and so the difference equations (\ref{eq:approx_B}) and (\ref{eq:approx_phi}) represent a {\it stochastic dynamical system}. By taking into account the equation (\ref{eq:approx_B}) the transmission through the chain of length $n$ can be written as a product
\beq
  B_n = B_1 \prod_{k=1}^n f(B(k),\phi_k + \alpha(k))\>.
  \label{eq:B_prod}
\eeq
In order to understand the scaling of the transmission with the length $n$ we introduce the {\it transmission decay rate} $I_n = \frac{1}{n}\log(B_n)$, which reads 
\beq
  I_n = \frac{1}{n} \sum_{k=1}^n \log [ f(B(k),\phi_k + \alpha(k))]\>, 
  \quad 
  n\gg 1\>.
\label{eq:In}  
\eeq  
We would like to express the distribution of $I_n$ over an ensemble of realisations of the chain in the limit $n \gg 1$, defined as
\beq
  P_n(I) = \ave{\delta(I - I_n)}_{\rm stoch.}\>,
  \label{eq:Pn}
\eeq
where $\ave{\ldots}_{\rm stoch.}$ denotes the average over $S\in{\cal A}$. To obtain this, we need to know the dynamics of the variable $\phi_n$ determined by (\ref{eq:approx_phi}). The distribution of the position of the dynamical system $\phi_n$ starting at point $\phi_1$ is defined by
\beq
  \rho_n(\phi; \phi_1) = \ave{\delta(\phi - \phi_n)}_{\rm stoch.}\>.
\eeq
It is meaningful to assume that $\rho_n$ has a limiting distribution independent of the initial position $\phi_1$ that is written as
\beq
  \rho(\phi) = \lim_{n\to\infty} \rho_n(\phi; \phi_1)\>.
  \label{eq:rho_phi}
\eeq
In some simple stochastic processes we can analytically express $\rho(\phi)$, but generally this is not the case. If  $\rho(\phi)$ exists then it is straightforward to show, using the central limit theorem \cite{feller:book:70}, that the limiting distribution of $I_n$ is a Gaussian distribution 
\beq
  P_n(I) = \sqrt{\frac{n}{2\pi\sigma_I^2}} 
  \exp\left(-n\frac{(I-\mean{I})^2}{2\sigma_I^2}\right)\>,\quad n\gg 1\>,
  \label{eq:gauss}
\eeq
with the first $\mean{I}$ and the second moment $\sigma^2_I$ given by
\beq
  \mean{I} = \ave{\log[f(B,\phi + \alpha]}\>,\qquad 
  \sigma_I^2 = \ave{\log[f(B,\phi + \alpha]^2} - \mean{I}^2\>,
  \label{eq:moments}
\eeq
where $\ave{\bullet}$ is an average over the stochastic variables $B$ and
$\alpha$, and over $\phi$ distributed with the density $\rho(\phi)$ (\ref{eq:rho_phi}). The result (\ref{eq:gauss}) implies that the transmission decays exponentially with the {\em normally} distributed rate (\ref{eq:gauss}). A similar result was already reported in \cite{abrahams:jpc:80, anderson:phys_rev:80} using scaling techniques and transfer matrix formalism, respectively, that are technically complicated compared to our derivation. The theoretical predictions are supported by numerical studies of which two examples are shown in figure \ref{pic:distr_I}. There is a good agreement between the measured distribution of $I_n$ and the one predicted theoretically.
\section{Chain of multi-channel scatterers}
We continue our discussion with the general linear chains of $d$-channel scatterers where the Lie group manifold of unitary $U(2d)$ scattering matrices is $(2d)^2$ dimensional \cite{elliot:book:79}. Large phase space dimension of lengthening dynamics makes analytical discussion very limited, hence our results that we report below are mostly numerical. The scattering matrices of the chain are again generated using the general recurrence relation (\ref{eq:S_map}). We limit ourselves to translationally invariant chains with fixed generating scattering matrices. Further, we restrict ourselves to the simplest case of sampling the S matrices from the entire unitary group $U(2d)$ with the corresponding Haar measure. For considering systems with time-reversal symmetry or some unitary (e.g. geometric symmetry), one could follow a similar approach but should first verify that the corresponding sub-space or sub-group of S matrices form a semi-group under the operation $\odot$ (\ref{eq:S_map}).
Here our discussion relies on the transfer matrices in the larger extent than in the single-channel case. Let us review some algebraic properties of the transfer matrices. The symmetry (\ref{eq:symm}) yields the following relations between eigenvalues and eigenvectors of the transfer matrix $T$
\beq
  Tv = \kappa v\quad \iff \quad (Kv)^\dag T = \frac{1}{\kappa^*} (Kv)^\dag\>,
\eeq
that for two eigenvalues $\kappa_{1,2}$ with the corresponding right eigenvectors $v_{1,2}$ the symmetry (\ref{eq:symm}) yields
\beq
  (\kappa_1\kappa_2^* - 1) v_2^\dag K v_1 = 0\>.
\eeq
This means that for every eigenvalue $\kappa$, with the right eigenvector $v$, there is a corresponding eigenvalue $1/\kappa^*$, with the left eigenvector $Kv$. In case when the eigenvalue lies on the unit circle, $|\kappa| = 1$, and is non-degenerate, the right eigenvector satisfies the relation $v^\dag K v \neq 0$. However, for the right eigenvector $v$ corresponding to the eigenvalue lying outside the unit circle we have $v^\dag Kv = 0$.\par
The dynamical system of scattering matrices $S_n$ defined by the recurrence (\ref{eq:S_map}) has all Lyapunov exponents \cite{reichl:book:04} equal to zero and so it is not chaotic. This becomes evident from the following discussion. Let us assume that we have a generating scattering matrix $S$ and the corresponding transfer matrix $T[S]$ with the spectrum $\{\kappa_1,\ldots, \kappa_{2d}\}$ and the corresponding right eigenvectors $v_i$. Then we can write the lengthening dynamics of the chain in the transfer matrix formalism as
\beq
  T_n = P\,\diag \{ e^{\kappa_i n}\}_{i=1}^{2d} P^{-1} \>.
  \label{eq:dyn_d}
\eeq
where the matrix $P$ has columns $v_i$, $P=[v_i]_{i=1}^{2d}$.
We distinguish two types of dynamics of the scattering matrix description of the chain depending on the spectral properties of the transfer matrix:\\
(i) If all eigenvalues lie on the unit circle, $|\kappa_i| = 1,\; \forall i$ the dynamics of $S_n$ is quasi-periodic.  This type of dynamics we call {\it ballistic motion} and the corresponding generating scattering matrices form a set of {\it ballistic unitary matrices} ${\cal M}_{\rm b}$.\\
(ii) If there is (at least one) eigenvalue outside the unit circle then the transmission ${\cal T}[S_n]$ of the chain decays exponentially towards some plateau value around which it oscillates as the chain is  lengthened. Following the definition of the transfer matrix (\ref{eq:ST_form}) and the relation with the scattering matrix (\ref{eq:ST_relation}) the average transmission probability can be expressed in terms of the lower-right diagonal block $X_n := (x_4)_n$ of the matrix $T_n$ as
\beq
  {\cal T}_n =\frac{1}{d} \tr \left\{ (X_n^\dag)^{-1} X_n^{-1} \right\}\>.
\eeq
The left eigenvectors of $T[S]$ are the rows in the inverse transition matrix $P^{-1} = [u^T_i]_{i=1}^{2d}$.  We write the right eigenvector as $v_i^T = [\alpha_i, \beta_i]\in\bC^{2d}$ and the left eigenvector as $u_i^T = [\zeta_i, \eta_i]\in\bC^{2d}$ by introducing the upper halves $\alpha_i,\zeta_i\in\bC^d$ and the lower halves $\beta_i,\eta_i\in\bC^d$ of eigenvectors. Then according to (\ref{eq:dyn_d}) the dynamics of the block $X_n$ reads
\beq
  X_n = \sum_{i=1}^{2d}  e^{\kappa_i n} \beta_i \eta_i^T\>.
\eeq
We denote by ${\cal K}=\{i:|\kappa_i|>1\}$ the set of indices corresponding to eigenvalues outside the unit circle. The vectors $u_i, v_i$ with $i\in{\cal K}$ satisfy the identity $v_i^\dag Kv_i = u_i^\dag Ku_i$, which implies that upper and lower halves of these vectors are non-trivial: $\|\alpha_i\|^2 = \|\beta_i\|^2\neq 0$ and  $\|\zeta_i\|^2 = \|\eta_i\|^2 \neq 0$. We introduce projection matrices $P_{\rm u}^\rR$ and $P_{\rm u}^\rL$ onto the set of vectors $\{\beta_i\}_{i\in {\cal K}}$ and $\{\eta_i\}_{i\in {\cal K}}$, respectively. Then the transmission can be expressed as a sum of non-decaying (oscillating) and decaying term:
\beqa
  {\cal T}[S_n] = {\cal T}_0[S_n]  + O(\exp(-\beta I n))\>,\qquad
  I = \log \max_m\{|\kappa_m|\}\>,
\eeqa
with $\beta\in\{1,2\}$, where the non-decaying term is written as
\beq
  {\cal T}_0[S_n] = 
  \tr \left\{ (\tilde X_n^\dag)^{-1}\tilde X_n^{-1}\right\}\>,\quad
  \tilde X_n = (\id-P_{\rm u}^\rL) X_{n} (\id-P_{\rm u}^\rR)\>, 
  \label{eq:non_decay}        
\eeq
where we introduce the decay rate $I$ analogous to that in the single-channel case. If the off-diagonal blocks of the matrix $X_n$ expressed in terms of introduced projectors are zero, $(\id-P_{\rm u}^\rL) X_n P_{\rm u}^\rR = P_{\rm u}^\rL X_n (\id - P_{\rm u}^\rR)=0$, then the coefficient $\beta=2$, otherwise $\beta=1$. The latter is statistically more likely situation. In case the block $X_n$ is approximately a random matrix then ${\cal T}_0 [S_n]\sim \#{\cal K}/d$, where $\#$ denotes the number of elements of a finite set. The presented dynamics of transmission is called {\it localised motion} and the set of corresponding generating scattering matrices are called {\it localised matrices} denoted by ${\cal M}_{\rm l}$. In case that all eigenvalues are out of the unit circle $\#{\cal K} = d$,  ($P^{\rR,\rL}_{\rm u}=\id$), the transmission decays to zero, and we are talking about {\it total localisation}, whereas the general case $1 \le \#{\cal K} \le d-1$ is referred to as {\it partial localisation}.\par
\begin{figure}[!htb]
  \centering
  \includegraphics[width=7.5cm]{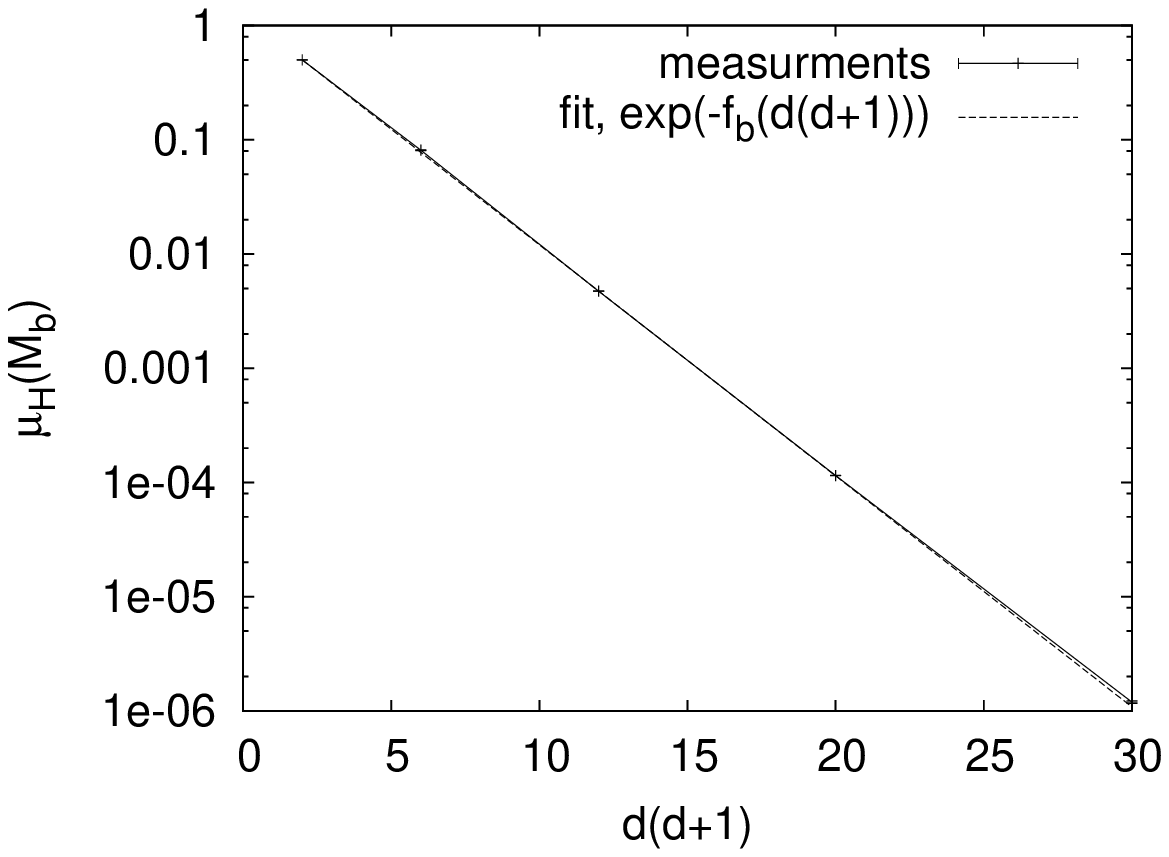}\hskip2pt%
  \includegraphics[width=8cm]{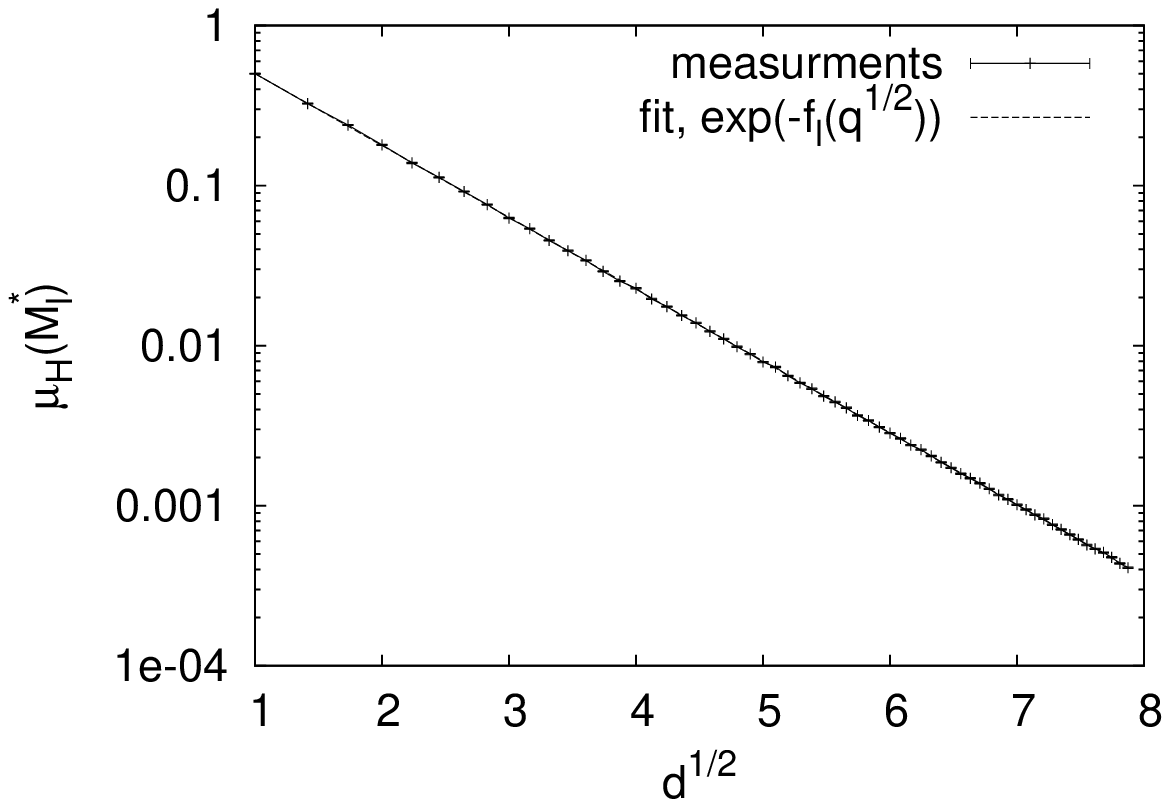}
  \hbox to16cm{\hfil (a) \hfil (b) \hfil}
\caption{The Haar measure of ballistic unitary matrices $\mu_\rH({\cal M}_{\rm b} \subset U(2d))$ (a), and totally localised unitary matrices $\mu_\rH({\cal M}_{\rm l}^* \subset U(2d))$ as a function of number of channels $d$. The fitted functions are $f_{\rm b}(x) = 0.4658x - 0.2387$ and $f_{\rm l}(x) = 1.034x - 0.342$ with $x=d(d+1)$ and $x=\sqrt{d}$, respectively.}
\label{pic:dim}
\end{figure}

The set of generating (unitary) scattering matrices is split into the set of ballistic ${\cal M}_{\rm b}$ and localised matrices ${\cal M}_{\rm l}$. An interesting subset of localised matrices are that corresponding to the total localisation. These matrices are named totally localised generating scattering matrices and their set is denoted by ${\cal M}_{\rm l}^*\subset {\cal M}_{\rm l}$. The separation between different sets of matrices is done on the ground of eigenvalues of the corresponding transfer matrices and it is interesting to know the Haar measures of these two sets. The measures are obtained numerically by generating unitary matrices uniformly with respect to the Haar measure \cite{reichl:book:04} and checking the spectrum of the corresponding transfer matrix for eigenvalues outside of unit circle. The result is plotted in figure \ref{pic:dim}.
We see that the measure of ballistic matrices decreases very fast with the channel number $d$. We can accurately fit numerical data with an empirical formula
\beq
  \mu_{\rm H}({\cal M}_{\rm b}) \approx \Omega \exp(- \omega d (d+1)), \quad 
  {\rm with}\quad \Omega \approx 0.7877,\;\omega\approx 0.4658\>.
\label{eq:Omega}
\eeq
In addition we observe that the measure of totally localised matrices also decreases with $d$, however slower, perhaps with a stretch-exponential law. Again we find very accurate empirical formula
\beq
  \mu_{\rm H}({\cal M}_{\rm l}^*) \approx Z\exp(-\zeta \sqrt{d}),\quad 
  {\rm with}\quad Z\approx 0.711,\;\zeta\approx 1.034\>.
\label{eq:Zeta}
\eeq
From these results we conclude that probability of partial localisation is quickly converging to $1$ as $d$ increases, however we have at present no theoretical explanation or derivation of (\ref{eq:Omega},\ref{eq:Zeta}). \par
\begin{figure}[!htb]
  \centering
  \includegraphics[width=8.7cm]{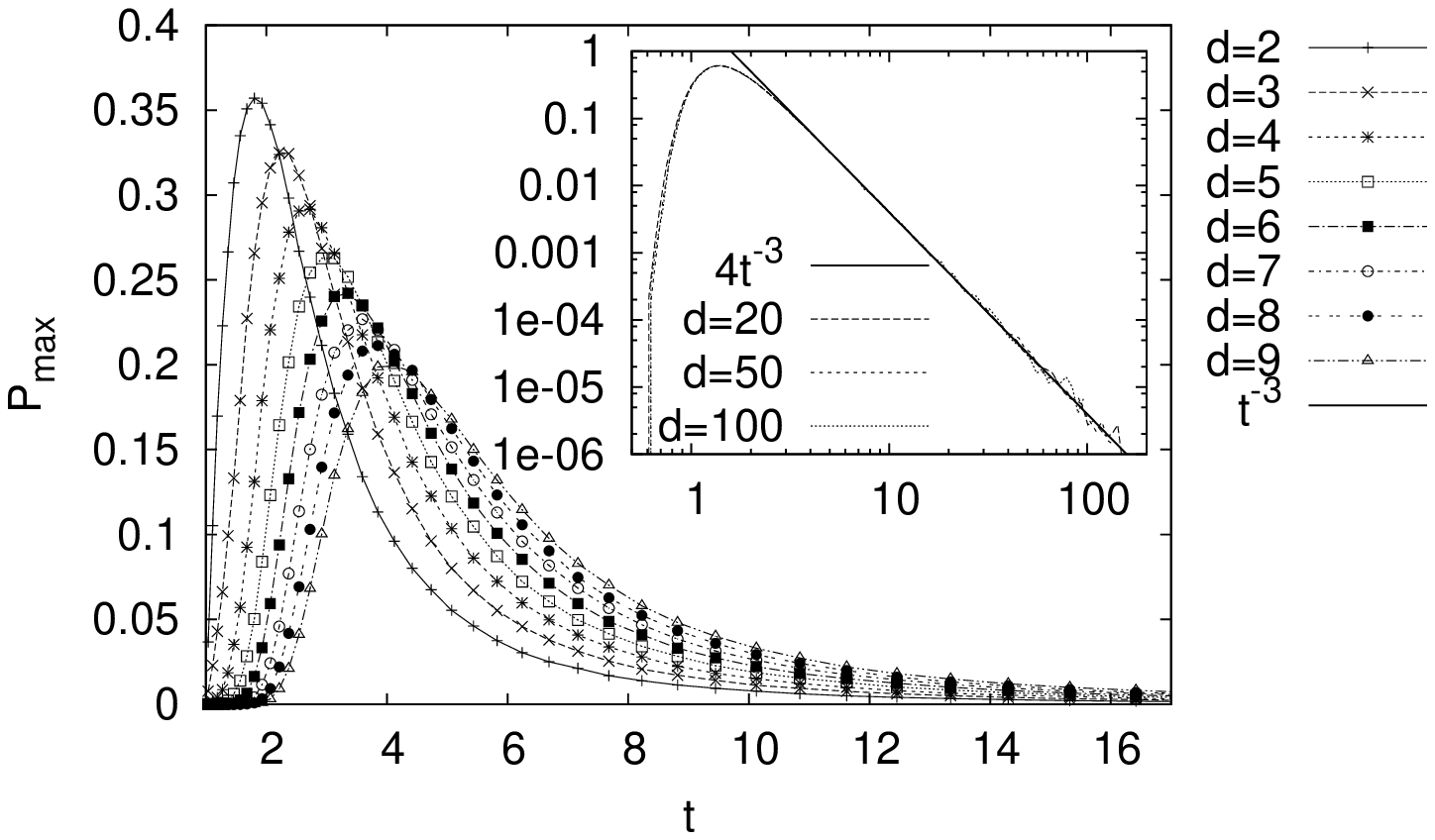}\hskip5pt%
  \includegraphics[width=6.8cm]{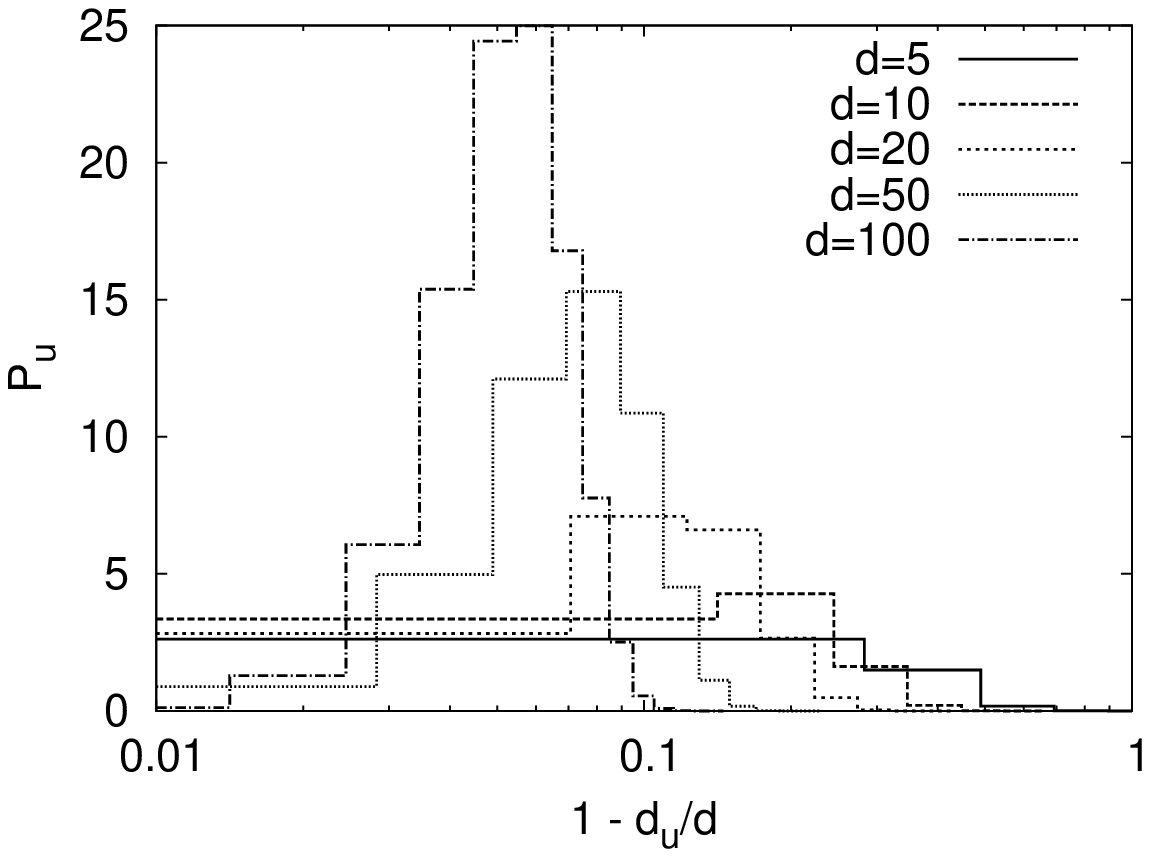}
  \hbox to 16cm{\hfil (a) \hfil (b) \hfil}
\caption{The distributions of maximal eigenvalue modulus
 of the transfer matrices $P_{\rm max}(t>1)$ (a), and the distribution of the relative number $x=d_{\rm u}/d$ of eigenvalues outside the unit circle $P_{\rm u}(x)$ plotted against $1-x$ (in log scale) (b). The inset in fig. (a) shows $\sqrt{d} P_{\rm max}(\sqrt{d} t)$ as a function of $t$ for several large $d$, in log-log scale.}
\label{pic:decay_charact}
\end{figure}

It is also instructive to study the eigenvalues of transfer matrices corresponding to localised scattering matrices ${\cal M}_{\rm l}$ as they give information about the length-scales of the transmission decay ${\cal T} [S_n]$. Let us write the spectrum of the transfer matrix $T[S]$ as 
\beq
  \Sigma[S]=\{\kappa: \det (T[S] - \kappa \id) = 0\}\>,
\eeq
and denote by $d_{\rm u}(S)$ the number of eigenvalues in $\Sigma[S]$ outside the unit circle. We investigate the distribution of the maximal eigenvalue modulus:
\beq
  P_{\rm max} (t; d) 
  = \int_{U(2d)} \dd \mu_\rH(S) \delta (t - \max|\Sigma[S]|)\>,\quad t\ge 1\>,
\eeq
and distribution of the relative number of eigenvalues outside the unit circle:
\beqa
  P_{\rm u} (t; d) 
  = \int_{U(2d)} \dd \mu_\rH(S) \delta (t - d_{\rm u}(S)/d)\>, \quad t \in [0,1]\>,
\eeqa
over the set of scattering matrices $S \in U(2d)$ with respect to the Haar measure $\mu_\rH$. From the distributions $P_{\rm max}$ and $P_{\rm u}$ we can learn about the decay rates and the percentage of unstable eigenvalues involved in the decay of transmission by lengthening of the chain, respectively. Both distributions are numerically calculated for several values of $d$ and shown in figure \ref{pic:decay_charact}. 
In figure \ref{pic:decay_charact}.a we see that $P_{\rm max}(t)$ has algebraic asymptotics, $P_{\rm max}(t;d) \sim t^{-3}$ as $t\to\infty$, and is zero on the unit circle $t = 1$. By closer inspection one can find that the distribution $P_{\rm max}(t)$ has an interesting semi-classical ($d \gg 1$) scaling property
\beq
  \lim_{d\to\infty} \sqrt{d} P_{\rm max}\left(\sqrt{d} t;d\right) = P(t)\>, 
  \qquad
  t > 0\>,
\eeq		
with $P(t)$ having asymptotic algebraic dependence
\beq
  P (t) \asymp a t^{-3},\qquad a \doteq 4.0 \pm 0.02\>,\quad t\to\infty\>.
\eeq
This scaling of $P_{\rm max}(t)$ implies that the average decay factor increases with the channel number $d$ as $t\sim\sqrt{d}$ and consequently the decay rate (inverse localisation length) increases as $I \sim \frac{1}{2}\log (d)$. The scaling law does not work for the single-channel case ($d=1$), but this is not very surprising. From results for the distribution of relative number of unstable directions $P_{\rm u}$, shown in \ref{pic:decay_charact}b, we see that the dimension $d_{\rm u}$ on average increases with increasing channel number $d$ and the average of the distribution moves towards the border value $d$. We conclude that on average the transmission for larger $d$ decays faster and to a lower asymptotic plateau given by $T_0[S_n]$ (\ref{eq:non_decay}).

\section{Summary and Conclusion}

We have devised an alternative strategy for the analysis of linear chains of scatterers. Our approach is based on defining the lengthening of the scattering chain as a dynamical system, and connecting its dynamical properties to physical (transport) properties of the chain.\par
The lengthening dynamics has been shown to be stable for arbitrary number of scattering channels. We have been able to reduce the single-channel case to a simple two-dimensional dynamical system in which we obtained many results analytically, for example the known regimes of ballistic or localised transport correspond to elliptic or attractive fixed point of lengthening dynamics, respectively.
In the general multi-channel case we were only able to give numerical results. We again separate the motion into ballistic and localised on the bases of the generating scattering matrices. In particular we give accurate numerical results on the scaling of Haar volumes of the sets of scattering matrices corresponding to totally localised, and ballistic transport. We have shown that in the regime of large number of scattering channels, the transport is most likely to be partially localised. In addition, we have examined the distributions of eigenvalues of transfer matrices with respect to the Haar measure for the corresponding scattering matrices. Interestingly, we found that the distribution of maximal eigenvalue modulus satisfies a scaling relation in the regime of large number of channels, corresponding for example to semiclassical situations (such as e.g. studied in \cite{horvat:phd:06}).
 
\section*{Acknowledgements}

Useful discussions with M. \v Znidari\v c, G. Veble and T. H. Seligman, as well as the financial support by the Ministry of higher education, science and technology of Slovenia are gratefully acknowledged.

\end{document}